\documentclass[prb,floats,eqsecnum,amsmath,amssymb]{revtex4}
\usepackage[dvips]{graphicx}

\begin{document}

\title {Direct coupling and inhomogeneity assist neurons to detect correlation in low amplitude noises}
\author {E. Bolhasani, Y. Azizi, and A. Valizadeh}

\affiliation{Institute for Advanced Studies in Basic Sciences, P.O. Box 45195--1159, Zanjan, Iran}

\begin{abstract}
We address a question on the effect of common stochastic inputs on the correlation of the spikes trains of two neurons when they are possibly nonidentical and are coupled through direct connections. We show that the change in the correlation of low amplitude stochastic inputs can be better detected when the neurons are connected by direct excitatory couplings. Depending on whether the neurons are identical or they are slightly different, symmetric or asymmetric connections can increase the sensitivity of the system to the input correlation by changing the mean slope of correlation transfer function over a given range of input correlation. In either case, there is also an optimum value for synaptic strength which maximizes the sensitivity of the system to the changes in input correlation.
\end{abstract}

\maketitle
\section{introduction}
The recent advent of novel recording techniques made it easier to simultaneously record from large number of neurons and provided new possibilities to relate population activity to coding and information processing in the brain \citealp{greenberg2008population, cohen2011measuring}. Many researchers suggest that studying the correlated activity of neurons in a population is essential for understanding how the information are coded in the brain \citealp{zohary1994correlated, abbott1999effect, nirenberg2003decoding, averbeck2006neural, biederlack2006brightness, schneidman2006weak, pillow2008spatio}. Correlated spiking of neurons contributes in several cognitive functions such as attention \citealp{steinmetz2000attention}, sensory coding \citealp{christopher1996primary, bair2001correlated, doiron2004oscillatory, schoppa2006synchronization, galan2006correlation} and discrimination \citealp{stopfer1997impaired, kenyon2004correlated}, motor behavior \citealp{maynard1999neuronal} and population coding \citealp{sompolinsky2001population, averbeck2006neural, josic2009stimulus}. Besides functional effects of such correlations between populations of neurons on neural coding, understanding how different parameters such as biological, network or stimulus parameters tune them is eventually being revealed \citealp{shadlen1998variable, binder2001relationship, moreno2002response, moreno2006auto, tchumatchenko2010correlations, rosenbaum2011membrane}. Correlation between neuronal activities is measured frequently by pairwise correlation coefficients and spike count correlations and ability of a neuronal system to transfer correlation can be quantified by correlation transfer function (CTF), which determines the relation between output correlation of a system under stimulus with specific input correlation \citealp{doiron2006stochastic, shea2008correlation, rosenbaum2011membrane}.


A periodic common input on two (or more) uncoupled oscillators can cause a coherent behavior when both the oscillators lock to the external force \citealp{pikovsky2003synchronization}. A very common example is the control of circadian rhythms of human/animals by the light-dark stimulation \citealp{roberts2005update}. In case of noisy inputs the counterpart of the phenomena appears as stochastic synchronization (SS) which is a general topic that addresses the phenomenon of irregular phase locking between two noisy non-linear oscillators \citealp{neiman1999synchronization}. In nervous systems, cross-correlations can arise either from the presence of direct synaptic connections \citealp{csicsvari1998reliability, bartho2004characterization}  or from shared inputs from the surrounding network or sensory layers \citealp{binder2001relationship, turker2001effects, turker2004effects}. While effect of direct synaptic connections and common inputs are vastly studied, less attention has been paid to the interplay of the two sources of correlation while they can be present concurrently in many physical and biological systems. We will show several nontrivial results can arise when two neurons with direct synaptic connections, are subjected to common/correlated inputs.

Possible differences between intrinsic parameters of neurons, causes the message from environment to the system to be decoded differently by the system individuals. Another aim of the current study is to investigate how the correlation is transferred by two neurons when the neurons are not identical. In such a heterogeneous system, the temporal symmetry of spike correlation is lost \citealp{tchumatchenko2010correlations}. We will show that even an slight inhomogeneity in the intrinsic parameters can lead to large reduction of pairwise correlation coefficient in the case of uncoupled neurons. As it is expectable the results depend on the time scales on which the correlation is calculated: Spike count correlations over long time bins are less affected by the heterogeneity but synchrony -- alignment of action potential in small time bins-- is tightly dependent on the homogeneity of the system.

  We have shown that correlated inputs and direct connections can either show cooperative or competing effects in different ranges of parameters. For uncoupled neurons, correlation susceptibility increases by increasing amplitude of noise for mildly correlated inputs \citealp{de2007correlation, shea2008correlation, tchumatchenko2010correlations}. Our results show that when direct connections are present between nonidentical neurons, the mean susceptibility is not anymore a monotonic function of amplitude of correlated noisy input. Reminiscence of stochastic resonance phenomena (SR), an intermediate noise amplitude in this case leads to larger sensitivity of the system to the changes in input correlation. We have also shown that with monosynaptic connections between two neurons, nonidentical neurons can show more correlation comparing to similar neurons. It means that with unidirectionally connections, slight inhomogeneity can increase the correlation of spike trains. Changing mismatch and synaptic strengths between two neurons, it is possible to change the functional form of correlation transfer function to optimize the mean susceptibility which is indicator of the sensitivity of the system to the change of input correlation in different ranges. In this way, as the most important result of current study, we will show that with direct couplings it is possible to detect correlation in small amplitude noises by increasing the sensitivity of the system to the change of correlation in the small amplitude noisy inputs.

\section{Material \& Methods}
The system under investigation consists of two coupled leaky integrate and fire (LIF) neurons \citealp{knight1972relationship}, subjected to correlated stochastic inputs. Subthreshold dynamics of the neuron in the LIF model obeys the following first order equation:
\begin{equation}\label{eq1}
    \tau_{m}\frac{dv_{i}}{dt}=V_{rest}-v_{i}+I_{i}+I_{ij},
\end{equation}
in which $v_{i}$ is a voltage-like variable for each neuron labeled by $i=1,2$ with $\tau_{m} = 20~\mathrm{ms}$ and $V_{rest} = -70~\mathrm{mV}$. A severe nonlinearity is imposed on the model by considering a threshold value $v_{th}=-54~\mathrm{mV}$. Whenever this value is reached, the neuron {\it spikes} and the voltage resets to $v_{reset}=-60~\mathrm{mV}$. (Parameters taken from \citealp{troyer1997physiological}). The spikes of the neurons are recorded as $x_{i}(t)=\sum_{m} \delta(t-t_{i}^{m})$ where $t_{i}^{m}$ is the time of $m^{th}$ spike of the neuron $i$, and $\delta(x)$ is the Dirac delta function.

Each model neuron receives a synaptic current through the direct connection from the other neuron $I_{ij}$, and an external current $I_{i}$ representing the sensory input or the effect of the surrounding networks. In the model equations, external current to the neuron $i$ comprises a constant (dc) and a stochastic component with amplitude $\sigma$. The stochastic inputs are sum of a common component $\xi_{c}(t)$ and an individual component $\xi_{i}(t)$:
\begin{equation}\label{eq2}
I_{i}(t)=(1\pm \delta)I + \sigma\left[\sqrt{1-c}\xi_{i}(t)+\sqrt{c}\xi_{c}(t)\right],
\end{equation}

\begin{figure}[htp]
\centerline{\includegraphics[width=25pc,angle=0]{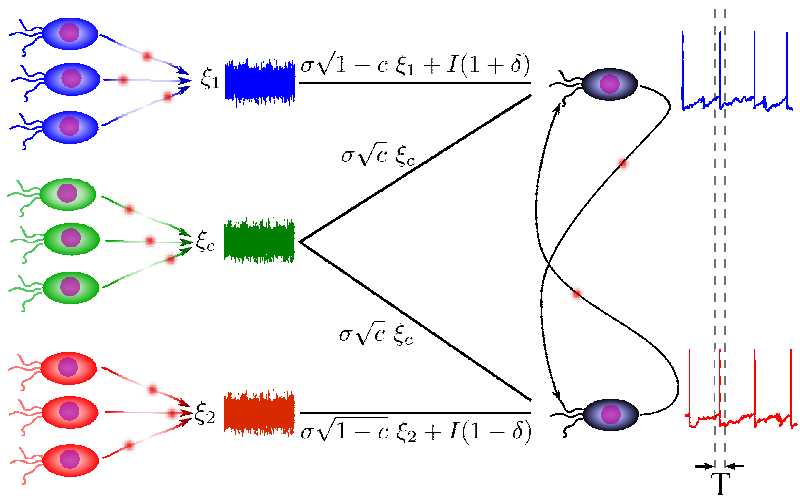}}
\caption{\textbf{Schematic representation of the model}. Two neurons stimulated by common and independent components, are possibly connected together by direct excitatory synaptic connections. Correlation of spike trains is then calculated over time bins much smaller than the mean inter-spike intervals.}\label{fig:1}
\end{figure}

where $\xi_{c}(t)$ and $\xi_{i}(t)$ are mutually independent Gaussian stochastic processes with zero mean and unit variance $\langle\xi_{i}(t)\xi_{j}(t')\rangle=\delta_{ij}\delta(t-t')$. The parameter $c \in [0,1]$ determines correlation of external currents which will be referred to as the input correlation. With the minimal model we used, inhomogeneity in the intrinsic activity rates is imposed by different constant currents which are chosen as $I_{1}=(1+\delta)I$ and $I_{2}=(1-\delta)I$, where $\delta$ is referred to as the parameter of inhomogeneity. With nonzero $\delta$ the neurons 1 and 2 will be the high frequency (fast) and low frequency (slow) neurons, respectively.

Neurons are pulse coupled. The neuron $i$ receives a pulse by the strength $\Delta_{ij}$ every time the neuron $j$ fires, so the synaptic current in Eq. \ref{eq1} can be written as $I_{ij}=\Delta_{ij} x_{j}(t)$ where the synaptic strength $\Delta_{ij}$ can be positive (excitatory) or negative (inhibitory). For convenience, we call the connections $21$ and $12$, the forward and backward connections, respectively. Note that although the external and synaptic inputs appear as currents, they are actually measured in units of the membrane potential $(\mathrm{mV})$ since a factor of the membrane resistance has been absorbed into their definition.

Co-fluctuations in the activity of neurons are measured over a range of timescales (for a review see \citealp{cohen2011measuring}). Spike count correlation is usually measured over the time scales from tens of milliseconds to seconds, while synchrony, that is almost precise alignment of the spikes, is measured over the time scale of the typical width of an action potential. It has been shown that spike count correlation over the small bins, bins of the order of one millisecond, can be largely determined by zero-lag conditional firing rate which quantifies exact synchrony \citealp{tchumatchenko2010signatures}. In this study we focus on synchrony, by describing spike counts and correlation coefficients in discrete bins of duration $T=0.5~\mathrm{ms}$ (unless otherwise noted). Correlation coefficient of spike counts $n_{i}(t)=\int_{t}^{t+T}x_{i}(s)ds$, is defined as the zero lag cross-correlation between $n_{1}$ and $n_{2}$:
\begin{equation}\label{eq3}
    \rho_{T}=\frac{\langle n_{1}(t) n_{2}(t)\rangle-\langle n_{1}(t) \rangle \langle n_{2}(t)\rangle}{\sqrt{\langle n_{1}(t)^2\rangle-\langle n_{1}(t) \rangle^2}\sqrt{\langle n_{2}(t)^2\rangle-\langle n_{2}(t) \rangle ^2}}.
\end{equation}

Correlation transfer function (CTF) is commonly used as a measure of dependence of the output correlation to the input correlation for a multi-element dynamical system \citealp{shea2008correlation, rosenbaum2011membrane}. To study sensitivity of correlation of output spike trains to the change of input correlation, we use \emph{differential correlation susceptibility} (DCS) as the mean slope of the correlation transfer function in  a given range of $c \in [c_1,c_2]$:
\begin{equation}\label{eq4}
    S_{T}(c_1,c_2)=\frac{\Delta \rho_{T}}{\Delta c}.
\end{equation}
which shows ratio of the change of correlation of spike trains $\Delta \rho_{T}= \rho_{T}(c_2)-\rho_{T}(c_1)$ to the change of input correlation $\Delta c=c_2-c_1$. For two identical neurons with no direct connection, this value is equal to one when it is evaluated over the full range of input correlation $[0,1]$.


\begin{figure*}[htp]
\centerline{\includegraphics[width=40pc,angle=0]{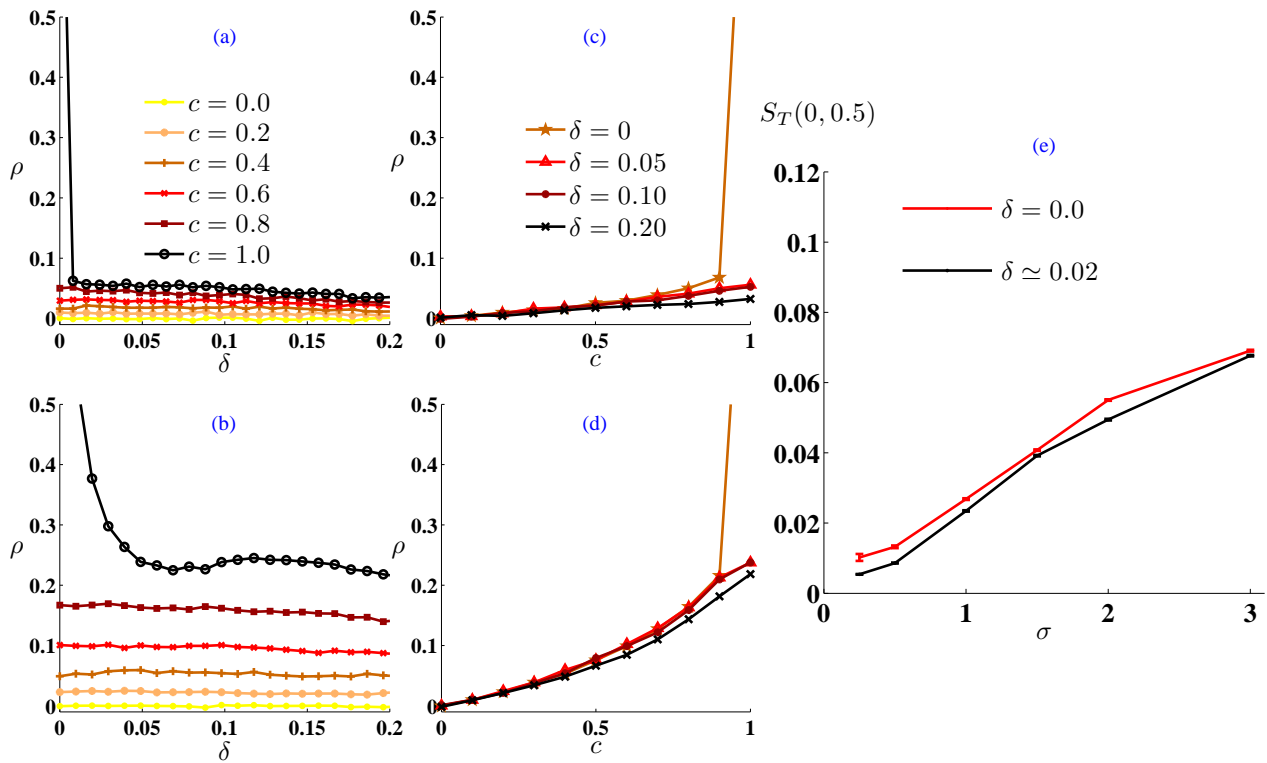}}
\caption{\textbf{Correlation of spike trains for two uncoupled neurons.} \textbf{(A)} Correlation coefficient is plotted against inhomogeneity, the mismatch between input current of neurons, for different values of input correlation. \textbf{(B)} Correlation transfer function CTF is plotted for different values of inhomogeneity. \textbf{(C,D)} The results are shown for a larger value of noise amplitude but with the same mean firing rate as \textbf{(A,B)} (see methods). \textbf{(E)} Differential correlation susceptibility (DCS) is plotted for homogeneous and slightly inhomogeneous system, as a function of noise amplitude, which shows the mean sensitivity of the output correlation to the change in input correlation over the range $[0,0.5]$.}\label{fig2}
\end{figure*}

\section{Results}
We first give the results for two uncoupled neurons. In Fig. 2\textbf{A} we have shown cross-correlation coefficient as a function of
mismatch between intrinsic firing rates of neurons for low noise amplitude and different values of input correlation. When there is no direct connection between the neurons, highly correlated inputs lead to large output correlation in case of identical neurons. Even a small mismatch decreases the output correlation considerably if the noise is low amplitude. In this case, even common noises lead to relatively low output correlations in presence of small inhomogeneity (e.g. $\delta=0.01)$ (Fig. 2\textbf{A)}. For larger noise amplitudes, the output correlation is less sensitive to inhomogeneity (Fig. 2\textbf{B}). The system is also less sensitive to inhomogeneity when the inputs are weakly correlated since both homogeneous and inhomogeneous systems have small output correlation. In Figs. 2\textbf{C} and 2\textbf{D} we have shown the correlation transfer function. It can be seen that while the slope of correlation transfer function decreases with mismatch for all the values of input correlation, this dependence is only considerable for highly (completely) correlated inputs. Increasing noise amplitude (while decreasing the constant input to avoid change in mean firing rate as explained below) makes the output correlation less sensitive to inhomogeneity for highly correlated inputs but yet the maximum sensitivity to the mismatch is seen for highly correlated inputs (Fig. 2\textbf{D}).

It has been previously shown that spike train correlation increases with firing rate \citealp{de2007correlation, shea2008correlation}. As noted above to avoid the results to be affected by the change in firing rate, we decreased mean value of input current while increasing the amplitude of fluctuations. The results shown in Figs. 2  and 3 are produced in such a way with a roughly constant mean value of firing rate $\sim 63Hz$. The mean value of the slope of the CTF over a given range of input correlations can quantify the average sensitivity of the spike train correlation to the correlation of inputs. In Fig. 2\textbf{E} we have plotted the DCS (mean slope of the correlation transfer function as described in methods) as a function of the amplitude of stochastic input for two uncoupled neurons over the range $c \in [0-0.5]$. The results for other ranges of partially correlated inputs are similar, i.e., the system shows low sensitivity to the change in input correlation for small amplitude noises and the sensitivity smoothly increases with increasing noise amplitudes. Just for the identical neurons if the full range of input correlation is considered ($c \in [0-1]$), the mean slope would be trivially equal to unity, independent of the noise amplitude.

These results show that the correlation between the low amplitude noises can not be suitably detected by a system of uncoupled neurons. To investigate the effect of direct couplings we have first considered a two neuron motif with just one unidirectional excitatory coupling. In many cases this configuration is favored when the synapses change through spike timing-dependent plasticity \citealp{song2000competitive}. We considered an excitatory forward coupling: From the high frequency (as the presynaptic) to low frequency neuron (as the postsynaptic) (see methods). In the absence of noise, any finite value of forward coupling strength can lead to a zone of $1:1$ synchrony, in which the dissimilar neurons fire in causal master-slave fashion \citealp{takahashi2009self, bayati2012effect}. In the causal limit the postsynaptic neuron fires immediately after receiving presynaptic stimulation \citealp{wang2012short, woodman2011effects}. In our model delays in communication have been ignored, so in causal $1:1$ synchrony zones, the postsynaptic neuron fires just one simulation time step after firing of presynaptic neuron. Since the time bin on which the correlation is calculated contains several time steps (see methods), such a causal master-slave firing leads to $\rho=1$ (Figs. 3\textbf{A} and 3\textbf{B} with no noise). Stochastic inputs have nontrivial effects on the correlation of the spike trains of these two neurons system. The output correlation is not anymore a monotonically decreasing function of mismatch, and namely a small mismatch can increase output correlation for correlated inputs (Figs. 3\textbf{A} and 3\textbf{B}).

\begin{figure*}[ht!]
\hspace{0cm}\centerline{\includegraphics[width=17cm,angle=0]{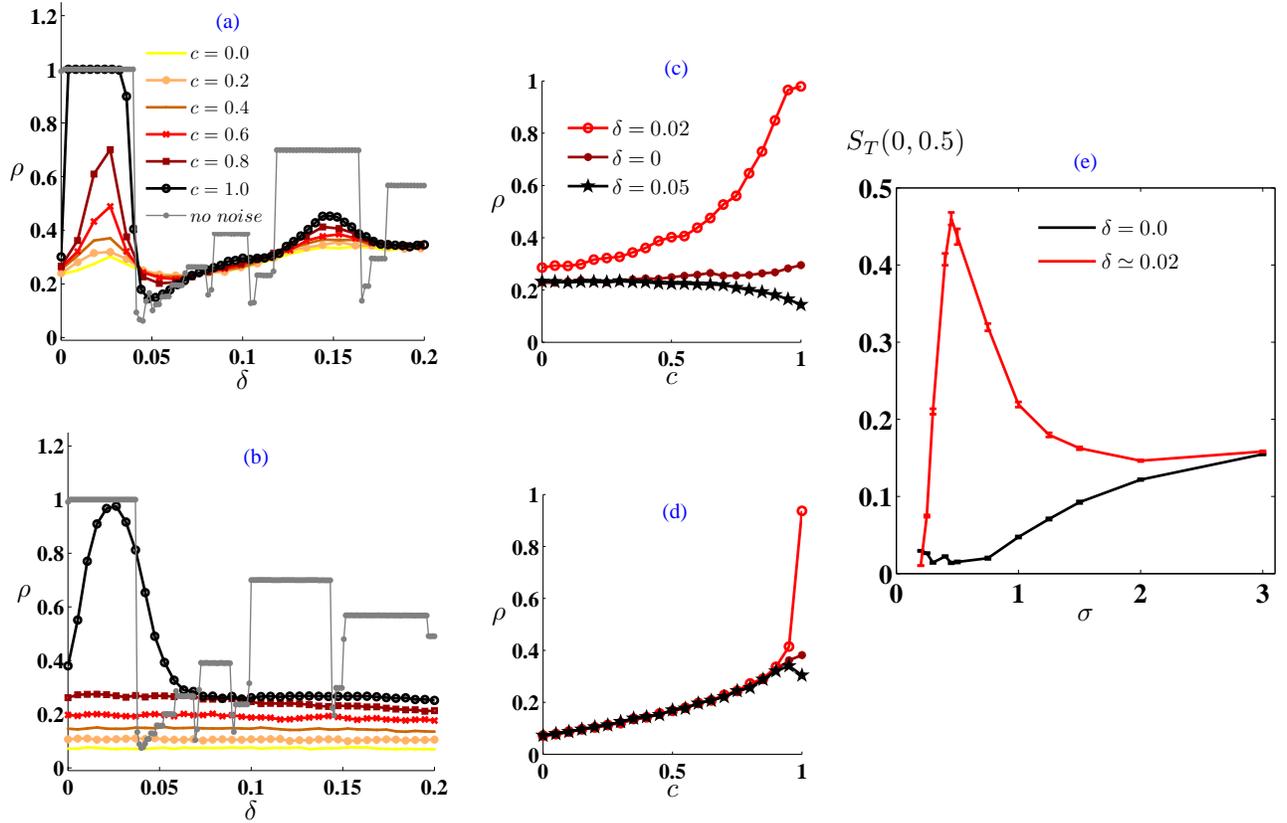}}
\vspace{0cm} \caption{\textbf{Correlation of spike trains for coupled neurons.} \textbf{(A-D)} All the results shown in Fig. 2 are repeated for two neurons, when the neurons are connected by a forward excitatory connection (from the high-frequency to the low-frequency neuron) of the strength $\Delta_{21}=1$.  The black curves in \textbf{(A)} and \textbf{(C)} correspond to autonomous case when no stochastic input is present.}\vspace{0cm} \label{fig3}
\end{figure*}

For coupled neurons, both the direct coupling and correlated inputs affect the correlation of spike trains and interesting effects can be seen when both the sources of correlation are present. Intuitively, the relative amplitudes of noise and recurrent stimulations determine the behavior of the system and the most notable results can be expected when the two sources are of the same order, i.e., when neither the external noise nor recurrent stimulations are dominant. In Figs. 3\textbf{C} and 3\textbf{D} we have shown CTF to inspect the effect of changing correlation of the stochastic inputs on correlation of spike trains for fixed value of synaptic strength. When the noise amplitude is not large, depending on the mismatch, different correlation transfer functions can be observed (Fig. 3\textbf{C}). Notably with changing mismatch it is possible to generate, for example, a CTF with higher sensitivity on the input correlation in different ranges of input correlation, or, a CTF with negative slope. Since firing rate can be changed by the mean synaptic inputs, it is possible that the coordinated response of the coupled neurons in a pool, to change over the time when their mean input change, say, in different levels of consciousness. In a different timescale, changes of synaptic strength through plasticity can also affect CTF for a given amplitude of noise.

As noted above, high level of stochastic input can wash out the effect of direct couplings and make the system to respond more similar to uncoupled neurons (see e.g. Fig. 3\textbf{D}). Impact of direct connections on the detection of input correlation of low amplitude noisy inputs is more apparent in a plot of DCS. In Fig. 4\textbf{A} we have plotted $S_T(0,0.5)$ as a function of noise amplitude for several values of synaptic strength, for unidirectionally coupled nonidentical neurons (with a small mismatch). Such an asymmetric (unidirectional) coupling has a crucial effect on DCS when it connects slightly different neurons (see also Fig. 4\textbf{D} and explanation below). As is shown in Fig. 4\textbf{A}, a forward monosynaptic connection (from high frequency to low frequency neuron) can considerably change the performance of the heterogeneous system in detecting variable input correlation. In an intermediate synaptic strength ($\Delta_{21}=1$) DCS shows faster growth in low amplitude noises and a higher maximum in relatively small amplitude noises. Further increasing of synaptic strength and noise amplitude reduces the performance of the system in detection of input correlation. With very large noise amplitudes, not presented in the figures, expectedly the effect of the direct connections is washed out and all the curves, including that of the uncoupled neurons, merge together and DCS smoothly increases with noise amplitude. Depending on the mismatch, there is an optimum value of synaptic strength which maximizes the mean sensitivity of correlation of spike trains to the input correlation. In Fig. 3\textbf{C} we have shown DCS as a function of the strength of forward unidirectional coupling for three values of mismatch. Expectedly, optimum value of synaptic strength is larger when the intrinsic firing rate of neurons are further different.

Overall increase of correlation of the spike trains is an intuitive expectation when direct couplings are present in the systems. But how direct couplings can increase the sensitivity on the changes in input correlation? Results shown in Fig. 3\textbf{B} indicate that the degree of amplification of output correlation depends on input correlation. A suitable choice of synaptic strength would result in more amplification for higher input correlations and increase the slope of correlation transfer function. Note that how very large synaptic strength decreases the sensitivity, due to the over-amplification of spike train correlation in small values of input correlation.

So far the results are shown when just a unidirectional excitatory coupling is present which is directed to low frequency neuron, and the neurons are nonidentical. To test other configurations, and to find the best configuration through which direct couplings can improve the performance of the system in detection of variable input correlation, we have tested mutual couplings with different ratios of forward $\Delta_{21}$ and backward $\Delta_{12}$ connections. While the sum of synaptic strengths are kept constant, different configurations can be designed by changing the ratio of coupling constants $r=\Delta_{21}/\Delta_{12}$ (Figs. 4\textbf{D} and 4\textbf{E}). In absence of mismatch, best configuration is that preserves symmetry, i.e., the best performance is resulted with equal forward and backward couplings. On the other hand, in presence of mismatch, an asymmetric arrangement of couplings in which the forward coupling (from the high frequency neuron) is larger, improves performance of the system. Interestingly, asymmetric excitatory couplings in favor of backward coupling (from the low frequency neuron), significantly decreases the sensitivity of the system.

\begin{figure*}[ht!]
\hspace{0cm}\centerline{\includegraphics[width=15cm,angle=0]{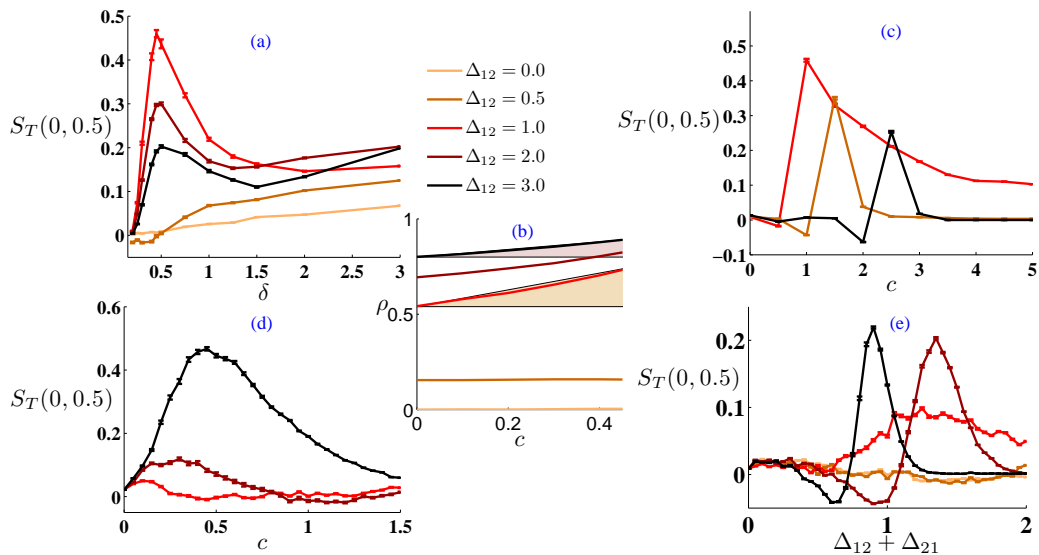}}
\vspace{0cm} \caption{\textbf{Differential correlation susceptibility.} \textbf{(A)} DCS is plotted vs. noise amplitude for two unidirectionally coupled nonidentical neurons ($\delta=...$). The results are shown for different values of synaptic strength. Maximum value of sensitivity to low amplitude noises can be obtained by an intermediate synaptic strength indicated in the figure. The dashed line shows the value of noise amplitude in (\textbf{C})-(\textbf{E}). \textbf{(B)} Correlation transfer function is plotted for different values of synaptic strength. Shadings are guide to eye for a comparison of the mean slope of the CTF for two different values of synaptic strength. \textbf{(C)} DCS is shown as a function of synaptic strength for different value of mismatch. The optimum value for synaptic strength grows for larger mismatch. \textbf{(D,E)} different configurations of couplings are tested for identical and nonidentical neurons, respectively. Different curves are plotted for different ratios $r$ of forward and backward couplings indicated in the legends. In \textbf{(D)} the neurons are identical and symmetric configuration $r=1$ shows the best performance with a suitable choice of synaptic strengths. In \textbf{(E)} nonidentical neurons have been tested: When the imbalance of couplings is in favor of forward coupling (from the high frequency neuron) the sensitivity considerably improves. When the backward coupling is larger $r<1$, the system performance is quite poor. As is shown in axes labels, DCS is calculated over the range $[0,0.5]$ of input correlation.}\vspace{0cm} \label{fig2}
\end{figure*}

\section{Discussion}

Both direct connections and common inputs can be source of correlation in the activity of neurons in nervous systems. While a large amount of literature has devoted to the effects of common inputs and direct connections \citealp{kuramoto1991collective, strogatz1991stability, abbott1993asynchronous}, less attention has been paid to the simultaneous effects of common inputs and direct connections \citealp{ostojic2009connectivity}. In this study we have numerically inspected effect of stochastic correlated inputs on the correlation of spike trains of two coupled LIF neurons. To study the system in a more general framework, we have considered the neurons with different intrinsic firing rates. We have assumed neurons with equal membrane time constants and inhomogeneity imposed on the system by feeding the neurons with different constant currents. The inhomogeneity, determined by the difference in the mean input currents, along with synaptic strengths are the key-parameters which specify the response of the system to the stochastic inputs.

 While for uncoupled neurons, output correlation is a monotonically decreasing function of inhomogeneity, for coupled neurons with low noise amplitudes, spike trains correlation can be increased by increasing inhomogeneity in some ranges. This result holds for sufficiently small noise amplitudes and the system inherits this property from $n:m$ locking zones for the autonomous system when there is no stochastic input present. This introduces inhomogeneity as an important parameter with nontrivial impact on the correlation of spike trains in coupled systems.

 Another feature of the system is that the two sources of correlation, correlated inputs and direct connections, do not necessarily cooperate in formation of correlated spike trains. Correlation transfer function determines the relation of output correlation to the input correlation, and its slope characterizes the sensitivity of the system to the change of input correlation. For uncoupled systems CTF is a monotonically increasing function and its slope decreases with lowering noise amplitude \citealp{de2007correlation, shea2008correlation} (if the inputs are not highly correlated) and with increasing mismatch (see Figs. 1\textbf{C} and \textbf{D} With different choices of synaptic strengths and inhomogeneity, it is possible to change CTF and design a system with different sensitivity to the input correlation. In particular, it is possible to design the system with negative mean slope of CTF, showing a case with destructive effect of common noises on the correlation of spike trains, and, the slope of CTF can be maximized in a range of input correlation. The latter proposes that direct connections can increase the sensitivity of the system to the correlation of the neuron's stochastic inputs, specially when the noises are low amplitude. We have further shown that for homogeneous system, with identical neurons, the best configuration of couplings which maximizes the mean sensitivity of the system in a given range, is a symmetric configuration, i.e., equal coupling constants. On the other hand, in presence of inhomogeneity, an asymmetric configuration in which the synaptic constant from the high frequency neuron to the low frequency is larger, improves sensitivity. In either case, there is an optimum value of synaptic constant which maximizes the sensitivity.

Competitive learning through classical spike timing-dependent plasticity (STDP) in feed-forward networks leads to potentiation of the synapses which convey correlated data and depression of those with uncorrelated activity \citealp{babadi2010intrinsic}. How STDP changes the lateral connections transverse to the path of data flow? It has been shown that in recurrent networks asymmetric connections arise through STDP and in presence of inhomogeneity, asymmetric change is in favor of the connection from the high frequency to low frequency neuron \citealp{takahashi2009self, bayati2012effect}. Our results show that asymmetric connections can enhance the performance of inhomogeneous systems in detection of input correlation, and interestingly such an optimum configuration of connections emerges through STDP (with asymmetric profile) in inhomogeneous neuronal pools \citealp{bayati2012effect}.

Type of neuronal excitability, shape of the phase resetting curves (PRCs), can also affect the correlation transfer in neuronal pools \citealp{galan2008optimal, abouzeid2009type, barreiro2010time}. Phase resetting curve characterizes how small perturbations influence the oscillator's subsequent timing or phase. It has been recently shown that type-II neurons with both negative and positive regions in their PRCs, transfer correlations more faithfully when the correlation is calculated over short time bins (\citealp{abouzeid2011correlation}). Although it needs further investigation but since the phase of a LIF neuron always advances in response to the external pulses, the results for LIF neurons are likely to apply for type-I neurons.

Correlation of spike trains over such small time bins we have used $T=0.5 ms$, is a measure of (almost) precise alignment of action potentials. Similar outputs resulted when we repeated the experiments with $T=1 ms$ but we expect qualitatively different results when correlation of spike counts is measured over the time scales, comparable, or larger than the mean inter-spike interval. Less sensitivity to inhomogeneity is expected when correlation is evaluated over large time bins, but effect of direct couplings warrants for further studies to find out if correlation in small amplitude stochastic inputs can be revealed in co-variation of spike trains of coupled neurons in large time scales.


\begin{references}
\expandafter\ifx\csname natexlab\endcsname\relax\def\natexlab#1{#1}\fi
\expandafter\ifx\csname bibnamefont\endcsname\relax
  \def\bibnamefont#1{#1}\fi
\expandafter\ifx\csname bibfnamefont\endcsname\relax
  \def\bibfnamefont#1{#1}\fi
\expandafter\ifx\csname citenamefont\endcsname\relax
  \def\citenamefont#1{#1}\fi
\expandafter\ifx\csname url\endcsname\relax
  \def\url#1{\texttt{#1}}\fi
\expandafter\ifx\csname urlprefix\endcsname\relax\def\urlprefix{URL }\fi
\providecommand{\bibinfo}[2]{#2}
\providecommand{\eprint}[2][]{\url{#2}}

\bibitem[{\citenamefont{Greenberg et~al.}(2008)\citenamefont{Greenberg,
  Houweling, and Kerr}}]{greenberg2008population}
\bibinfo{author}{\bibfnamefont{D.~S.} \bibnamefont{Greenberg}},
  \bibinfo{author}{\bibfnamefont{A.~R.} \bibnamefont{Houweling}},
  \bibnamefont{and} \bibinfo{author}{\bibfnamefont{J.~N.} \bibnamefont{Kerr}},
  \bibinfo{journal}{Nature neuroscience} \textbf{\bibinfo{volume}{11}},
  \bibinfo{pages}{749} (\bibinfo{year}{2008}).

\bibitem[{\citenamefont{Cohen and Kohn}(2011)}]{cohen2011measuring}
\bibinfo{author}{\bibfnamefont{M.~R.} \bibnamefont{Cohen}} \bibnamefont{and}
  \bibinfo{author}{\bibfnamefont{A.}~\bibnamefont{Kohn}},
  \bibinfo{journal}{Nature neuroscience} \textbf{\bibinfo{volume}{14}},
  \bibinfo{pages}{811} (\bibinfo{year}{2011}).

\bibitem[{\citenamefont{Zohary et~al.}(1994)\citenamefont{Zohary, Shadlen, and
  Newsome}}]{zohary1994correlated}
\bibinfo{author}{\bibfnamefont{E.}~\bibnamefont{Zohary}},
  \bibinfo{author}{\bibfnamefont{M.~N.} \bibnamefont{Shadlen}},
  \bibnamefont{and} \bibinfo{author}{\bibfnamefont{W.~T.}
  \bibnamefont{Newsome}}, \bibinfo{journal}{Nature}
  \textbf{\bibinfo{volume}{370}}, \bibinfo{pages}{140} (\bibinfo{year}{1994}).

\bibitem[{\citenamefont{Abbott and Dayan}(1999)}]{abbott1999effect}
\bibinfo{author}{\bibfnamefont{L.}~\bibnamefont{Abbott}} \bibnamefont{and}
  \bibinfo{author}{\bibfnamefont{P.}~\bibnamefont{Dayan}},
  \bibinfo{journal}{Neural Computation} \textbf{\bibinfo{volume}{11}},
  \bibinfo{pages}{91} (\bibinfo{year}{1999}).

\bibitem[{\citenamefont{Nirenberg and Latham}(2003)}]{nirenberg2003decoding}
\bibinfo{author}{\bibfnamefont{S.}~\bibnamefont{Nirenberg}} \bibnamefont{and}
  \bibinfo{author}{\bibfnamefont{P.~E.} \bibnamefont{Latham}},
  \bibinfo{journal}{Proceedings of the National Academy of Sciences}
  \textbf{\bibinfo{volume}{100}}, \bibinfo{pages}{7348} (\bibinfo{year}{2003}).

\bibitem[{\citenamefont{Averbeck et~al.}(2006)\citenamefont{Averbeck, Latham,
  and Pouget}}]{averbeck2006neural}
\bibinfo{author}{\bibfnamefont{B.~B.} \bibnamefont{Averbeck}},
  \bibinfo{author}{\bibfnamefont{P.~E.} \bibnamefont{Latham}},
  \bibnamefont{and} \bibinfo{author}{\bibfnamefont{A.}~\bibnamefont{Pouget}},
  \bibinfo{journal}{Nature Reviews Neuroscience} \textbf{\bibinfo{volume}{7}},
  \bibinfo{pages}{358} (\bibinfo{year}{2006}).

\bibitem[{\citenamefont{Biederlack et~al.}(2006)\citenamefont{Biederlack,
  Castelo-Branco, Neuenschwander, Wheeler, Singer, and
  Nikoli{\'c}}}]{biederlack2006brightness}
\bibinfo{author}{\bibfnamefont{J.}~\bibnamefont{Biederlack}},
  \bibinfo{author}{\bibfnamefont{M.}~\bibnamefont{Castelo-Branco}},
  \bibinfo{author}{\bibfnamefont{S.}~\bibnamefont{Neuenschwander}},
  \bibinfo{author}{\bibfnamefont{D.~W.} \bibnamefont{Wheeler}},
  \bibinfo{author}{\bibfnamefont{W.}~\bibnamefont{Singer}}, \bibnamefont{and}
  \bibinfo{author}{\bibfnamefont{D.}~\bibnamefont{Nikoli{\'c}}},
  \bibinfo{journal}{Neuron} \textbf{\bibinfo{volume}{52}},
  \bibinfo{pages}{1073} (\bibinfo{year}{2006}).

\bibitem[{\citenamefont{Schneidman et~al.}(2006)\citenamefont{Schneidman,
  Berry, Segev, and Bialek}}]{schneidman2006weak}
\bibinfo{author}{\bibfnamefont{E.}~\bibnamefont{Schneidman}},
  \bibinfo{author}{\bibfnamefont{M.~J.} \bibnamefont{Berry}},
  \bibinfo{author}{\bibfnamefont{R.}~\bibnamefont{Segev}}, \bibnamefont{and}
  \bibinfo{author}{\bibfnamefont{W.}~\bibnamefont{Bialek}},
  \bibinfo{journal}{Nature} \textbf{\bibinfo{volume}{440}},
  \bibinfo{pages}{1007} (\bibinfo{year}{2006}).

\bibitem[{\citenamefont{Pillow et~al.}(2008)\citenamefont{Pillow, Shlens,
  Paninski, Sher, Litke, Chichilnisky, and Simoncelli}}]{pillow2008spatio}
\bibinfo{author}{\bibfnamefont{J.~W.} \bibnamefont{Pillow}},
  \bibinfo{author}{\bibfnamefont{J.}~\bibnamefont{Shlens}},
  \bibinfo{author}{\bibfnamefont{L.}~\bibnamefont{Paninski}},
  \bibinfo{author}{\bibfnamefont{A.}~\bibnamefont{Sher}},
  \bibinfo{author}{\bibfnamefont{A.~M.} \bibnamefont{Litke}},
  \bibinfo{author}{\bibfnamefont{E.}~\bibnamefont{Chichilnisky}},
  \bibnamefont{and} \bibinfo{author}{\bibfnamefont{E.~P.}
  \bibnamefont{Simoncelli}}, \bibinfo{journal}{Nature}
  \textbf{\bibinfo{volume}{454}}, \bibinfo{pages}{995} (\bibinfo{year}{2008}).

\bibitem[{\citenamefont{Steinmetz et~al.}(2000)\citenamefont{Steinmetz, Roy,
  Fitzgerald, Hsiao, Johnson, and Niebur}}]{steinmetz2000attention}
\bibinfo{author}{\bibfnamefont{P.~N.} \bibnamefont{Steinmetz}},
  \bibinfo{author}{\bibfnamefont{A.}~\bibnamefont{Roy}},
  \bibinfo{author}{\bibfnamefont{P.}~\bibnamefont{Fitzgerald}},
  \bibinfo{author}{\bibfnamefont{S.}~\bibnamefont{Hsiao}},
  \bibinfo{author}{\bibfnamefont{K.}~\bibnamefont{Johnson}}, \bibnamefont{and}
  \bibinfo{author}{\bibfnamefont{E.}~\bibnamefont{Niebur}},
  \bibinfo{journal}{Nature} \textbf{\bibinfo{volume}{404}},
  \bibinfo{pages}{131} (\bibinfo{year}{2000}).

\bibitem[{\citenamefont{Christopher~deCharms and
  Merzenich}(1996)}]{christopher1996primary}
\bibinfo{author}{\bibfnamefont{R.}~\bibnamefont{Christopher~deCharms}}
  \bibnamefont{and} \bibinfo{author}{\bibfnamefont{M.~M.}
  \bibnamefont{Merzenich}}, \bibinfo{journal}{Nature}
  \textbf{\bibinfo{volume}{381}}, \bibinfo{pages}{13} (\bibinfo{year}{1996}).

\bibitem[{\citenamefont{Bair et~al.}(2001)\citenamefont{Bair, Zohary, and
  Newsome}}]{bair2001correlated}
\bibinfo{author}{\bibfnamefont{W.}~\bibnamefont{Bair}},
  \bibinfo{author}{\bibfnamefont{E.}~\bibnamefont{Zohary}}, \bibnamefont{and}
  \bibinfo{author}{\bibfnamefont{W.~T.} \bibnamefont{Newsome}},
  \bibinfo{journal}{The journal of Neuroscience} \textbf{\bibinfo{volume}{21}},
  \bibinfo{pages}{1676} (\bibinfo{year}{2001}).

\bibitem[{\citenamefont{Doiron et~al.}(2004)\citenamefont{Doiron, Lindner,
  Longtin, Maler, and Bastian}}]{doiron2004oscillatory}
\bibinfo{author}{\bibfnamefont{B.}~\bibnamefont{Doiron}},
  \bibinfo{author}{\bibfnamefont{B.}~\bibnamefont{Lindner}},
  \bibinfo{author}{\bibfnamefont{A.}~\bibnamefont{Longtin}},
  \bibinfo{author}{\bibfnamefont{L.}~\bibnamefont{Maler}}, \bibnamefont{and}
  \bibinfo{author}{\bibfnamefont{J.}~\bibnamefont{Bastian}},
  \bibinfo{journal}{Physical review letters} \textbf{\bibinfo{volume}{93}},
  \bibinfo{pages}{48101} (\bibinfo{year}{2004}).

\bibitem[{\citenamefont{Schoppa}(2006)}]{schoppa2006synchronization}
\bibinfo{author}{\bibfnamefont{N.~E.} \bibnamefont{Schoppa}},
  \bibinfo{journal}{Neuron} \textbf{\bibinfo{volume}{49}}, \bibinfo{pages}{271}
  (\bibinfo{year}{2006}).

\bibitem[{\citenamefont{Gal{\'a}n et~al.}(2006)\citenamefont{Gal{\'a}n,
  Fourcaud-Trocm{\'e}, Ermentrout, and Urban}}]{galan2006correlation}
\bibinfo{author}{\bibfnamefont{R.~F.} \bibnamefont{Gal{\'a}n}},
  \bibinfo{author}{\bibfnamefont{N.}~\bibnamefont{Fourcaud-Trocm{\'e}}},
  \bibinfo{author}{\bibfnamefont{G.~B.} \bibnamefont{Ermentrout}},
  \bibnamefont{and} \bibinfo{author}{\bibfnamefont{N.~N.} \bibnamefont{Urban}},
  \bibinfo{journal}{The Journal of neuroscience} \textbf{\bibinfo{volume}{26}},
  \bibinfo{pages}{3646} (\bibinfo{year}{2006}).

\bibitem[{\citenamefont{Stopfer et~al.}(1997)\citenamefont{Stopfer, Bhagavan,
  Smith, Laurent et~al.}}]{stopfer1997impaired}
\bibinfo{author}{\bibfnamefont{M.}~\bibnamefont{Stopfer}},
  \bibinfo{author}{\bibfnamefont{S.}~\bibnamefont{Bhagavan}},
  \bibinfo{author}{\bibfnamefont{B.~H.} \bibnamefont{Smith}},
  \bibinfo{author}{\bibfnamefont{G.}~\bibnamefont{Laurent}},
  \bibnamefont{et~al.}, \bibinfo{journal}{Nature}
  \textbf{\bibinfo{volume}{390}}, \bibinfo{pages}{70} (\bibinfo{year}{1997}).

\bibitem[{\citenamefont{Kenyon et~al.}(2004)\citenamefont{Kenyon, Theiler,
  George, Travis, and Marshak}}]{kenyon2004correlated}
\bibinfo{author}{\bibfnamefont{G.~T.} \bibnamefont{Kenyon}},
  \bibinfo{author}{\bibfnamefont{J.}~\bibnamefont{Theiler}},
  \bibinfo{author}{\bibfnamefont{J.~S.} \bibnamefont{George}},
  \bibinfo{author}{\bibfnamefont{B.~J.} \bibnamefont{Travis}},
  \bibnamefont{and} \bibinfo{author}{\bibfnamefont{D.~W.}
  \bibnamefont{Marshak}}, \bibinfo{journal}{Neural computation}
  \textbf{\bibinfo{volume}{16}}, \bibinfo{pages}{2261} (\bibinfo{year}{2004}).

\bibitem[{\citenamefont{Maynard et~al.}(1999)\citenamefont{Maynard,
  Hatsopoulos, Ojakangas, Acuna, Sanes, Normann, and
  Donoghue}}]{maynard1999neuronal}
\bibinfo{author}{\bibfnamefont{E.}~\bibnamefont{Maynard}},
  \bibinfo{author}{\bibfnamefont{N.}~\bibnamefont{Hatsopoulos}},
  \bibinfo{author}{\bibfnamefont{C.}~\bibnamefont{Ojakangas}},
  \bibinfo{author}{\bibfnamefont{B.}~\bibnamefont{Acuna}},
  \bibinfo{author}{\bibfnamefont{J.}~\bibnamefont{Sanes}},
  \bibinfo{author}{\bibfnamefont{R.}~\bibnamefont{Normann}}, \bibnamefont{and}
  \bibinfo{author}{\bibfnamefont{J.}~\bibnamefont{Donoghue}},
  \bibinfo{journal}{The journal of Neuroscience} \textbf{\bibinfo{volume}{19}},
  \bibinfo{pages}{8083} (\bibinfo{year}{1999}).

\bibitem[{\citenamefont{Sompolinsky et~al.}(2001)\citenamefont{Sompolinsky,
  Yoon, Kang, and Shamir}}]{sompolinsky2001population}
\bibinfo{author}{\bibfnamefont{H.}~\bibnamefont{Sompolinsky}},
  \bibinfo{author}{\bibfnamefont{H.}~\bibnamefont{Yoon}},
  \bibinfo{author}{\bibfnamefont{K.}~\bibnamefont{Kang}}, \bibnamefont{and}
  \bibinfo{author}{\bibfnamefont{M.}~\bibnamefont{Shamir}},
  \bibinfo{journal}{Physical Review E} \textbf{\bibinfo{volume}{64}},
  \bibinfo{pages}{051904} (\bibinfo{year}{2001}).

\bibitem[{\citenamefont{Josic et~al.}(2009)\citenamefont{Josic, Shea-Brown,
  Doiron, and De~La~Rocha}}]{josic2009stimulus}
\bibinfo{author}{\bibfnamefont{K.}~\bibnamefont{Josic}},
  \bibinfo{author}{\bibfnamefont{E.}~\bibnamefont{Shea-Brown}},
  \bibinfo{author}{\bibfnamefont{B.}~\bibnamefont{Doiron}}, \bibnamefont{and}
  \bibinfo{author}{\bibfnamefont{J.}~\bibnamefont{De~La~Rocha}},
  \bibinfo{journal}{Neural computation} \textbf{\bibinfo{volume}{21}},
  \bibinfo{pages}{2774} (\bibinfo{year}{2009}).

\bibitem[{\citenamefont{Shadlen and Newsome}(1998)}]{shadlen1998variable}
\bibinfo{author}{\bibfnamefont{M.~N.} \bibnamefont{Shadlen}} \bibnamefont{and}
  \bibinfo{author}{\bibfnamefont{W.~T.} \bibnamefont{Newsome}},
  \bibinfo{journal}{The Journal of Neuroscience} \textbf{\bibinfo{volume}{18}},
  \bibinfo{pages}{3870} (\bibinfo{year}{1998}).

\bibitem[{\citenamefont{Binder and Powers}(2001)}]{binder2001relationship}
\bibinfo{author}{\bibfnamefont{M.~D.} \bibnamefont{Binder}} \bibnamefont{and}
  \bibinfo{author}{\bibfnamefont{R.~K.} \bibnamefont{Powers}},
  \bibinfo{journal}{Journal of Neurophysiology} \textbf{\bibinfo{volume}{86}},
  \bibinfo{pages}{2266} (\bibinfo{year}{2001}).

\bibitem[{\citenamefont{Moreno et~al.}(2002)\citenamefont{Moreno, de~La~Rocha,
  Renart, and Parga}}]{moreno2002response}
\bibinfo{author}{\bibfnamefont{R.}~\bibnamefont{Moreno}},
  \bibinfo{author}{\bibfnamefont{J.}~\bibnamefont{de~La~Rocha}},
  \bibinfo{author}{\bibfnamefont{A.}~\bibnamefont{Renart}}, \bibnamefont{and}
  \bibinfo{author}{\bibfnamefont{N.}~\bibnamefont{Parga}},
  \bibinfo{journal}{Physical review letters} \textbf{\bibinfo{volume}{89}},
  \bibinfo{pages}{288101} (\bibinfo{year}{2002}).

\bibitem[{\citenamefont{Moreno-Bote and Parga}(2006)}]{moreno2006auto}
\bibinfo{author}{\bibfnamefont{R.}~\bibnamefont{Moreno-Bote}} \bibnamefont{and}
  \bibinfo{author}{\bibfnamefont{N.}~\bibnamefont{Parga}},
  \bibinfo{journal}{Physical review letters} \textbf{\bibinfo{volume}{96}},
  \bibinfo{pages}{28101} (\bibinfo{year}{2006}).

\bibitem[{\citenamefont{Tchumatchenko
  et~al.}(2010{\natexlab{a}})\citenamefont{Tchumatchenko, Malyshev, Geisel,
  Volgushev, and Wolf}}]{tchumatchenko2010correlations}
\bibinfo{author}{\bibfnamefont{T.}~\bibnamefont{Tchumatchenko}},
  \bibinfo{author}{\bibfnamefont{A.}~\bibnamefont{Malyshev}},
  \bibinfo{author}{\bibfnamefont{T.}~\bibnamefont{Geisel}},
  \bibinfo{author}{\bibfnamefont{M.}~\bibnamefont{Volgushev}},
  \bibnamefont{and} \bibinfo{author}{\bibfnamefont{F.}~\bibnamefont{Wolf}},
  \bibinfo{journal}{Physical review letters} \textbf{\bibinfo{volume}{104}},
  \bibinfo{pages}{58102} (\bibinfo{year}{2010}{\natexlab{a}}).

\bibitem[{\citenamefont{Rosenbaum and Josi{\'c}}(2011)}]{rosenbaum2011membrane}
\bibinfo{author}{\bibfnamefont{R.}~\bibnamefont{Rosenbaum}} \bibnamefont{and}
  \bibinfo{author}{\bibfnamefont{K.}~\bibnamefont{Josi{\'c}}},
  \bibinfo{journal}{Physical Review E} \textbf{\bibinfo{volume}{84}},
  \bibinfo{pages}{051902} (\bibinfo{year}{2011}).

\bibitem[{\citenamefont{Doiron et~al.}(2006)\citenamefont{Doiron, Rinzel, and
  Reyes}}]{doiron2006stochastic}
\bibinfo{author}{\bibfnamefont{B.}~\bibnamefont{Doiron}},
  \bibinfo{author}{\bibfnamefont{J.}~\bibnamefont{Rinzel}}, \bibnamefont{and}
  \bibinfo{author}{\bibfnamefont{A.}~\bibnamefont{Reyes}},
  \bibinfo{journal}{Physical Review E} \textbf{\bibinfo{volume}{74}},
  \bibinfo{pages}{030903} (\bibinfo{year}{2006}).

\bibitem[{\citenamefont{Shea-Brown et~al.}(2008)\citenamefont{Shea-Brown,
  Josi{\'c}, de~La~Rocha, and Doiron}}]{shea2008correlation}
\bibinfo{author}{\bibfnamefont{E.}~\bibnamefont{Shea-Brown}},
  \bibinfo{author}{\bibfnamefont{K.}~\bibnamefont{Josi{\'c}}},
  \bibinfo{author}{\bibfnamefont{J.}~\bibnamefont{de~La~Rocha}},
  \bibnamefont{and} \bibinfo{author}{\bibfnamefont{B.}~\bibnamefont{Doiron}},
  \bibinfo{journal}{Physical review letters} \textbf{\bibinfo{volume}{100}},
  \bibinfo{pages}{108102} (\bibinfo{year}{2008}).

\bibitem[{\citenamefont{Pikovsky et~al.}(2003)\citenamefont{Pikovsky,
  Rosenblum, and Kurths}}]{pikovsky2003synchronization}
\bibinfo{author}{\bibfnamefont{A.}~\bibnamefont{Pikovsky}},
  \bibinfo{author}{\bibfnamefont{M.}~\bibnamefont{Rosenblum}},
  \bibnamefont{and} \bibinfo{author}{\bibfnamefont{J.}~\bibnamefont{Kurths}},
  \emph{\bibinfo{title}{Synchronization: A universal concept in nonlinear
  sciences}}, vol.~\bibinfo{volume}{12} (\bibinfo{publisher}{Cambridge
  university press}, \bibinfo{year}{2003}).

\bibitem[{\citenamefont{Roberts}(2005)}]{roberts2005update}
\bibinfo{author}{\bibfnamefont{J.~E.} \bibnamefont{Roberts}},
  \bibinfo{journal}{Photochemistry and photobiology}
  \textbf{\bibinfo{volume}{81}}, \bibinfo{pages}{490} (\bibinfo{year}{2005}).

\bibitem[{\citenamefont{Neiman et~al.}(1999)\citenamefont{Neiman,
  Schimansky-Geier, Moss, Shulgin, and Collins}}]{neiman1999synchronization}
\bibinfo{author}{\bibfnamefont{A.}~\bibnamefont{Neiman}},
  \bibinfo{author}{\bibfnamefont{L.}~\bibnamefont{Schimansky-Geier}},
  \bibinfo{author}{\bibfnamefont{F.}~\bibnamefont{Moss}},
  \bibinfo{author}{\bibfnamefont{B.}~\bibnamefont{Shulgin}}, \bibnamefont{and}
  \bibinfo{author}{\bibfnamefont{J.~J.} \bibnamefont{Collins}},
  \bibinfo{journal}{Physical Review E} \textbf{\bibinfo{volume}{60}},
  \bibinfo{pages}{284} (\bibinfo{year}{1999}).

\bibitem[{\citenamefont{Csicsvari et~al.}(1998)\citenamefont{Csicsvari, Hirase,
  Czurko, and Buzs{\'a}ki}}]{csicsvari1998reliability}
\bibinfo{author}{\bibfnamefont{J.}~\bibnamefont{Csicsvari}},
  \bibinfo{author}{\bibfnamefont{H.}~\bibnamefont{Hirase}},
  \bibinfo{author}{\bibfnamefont{A.}~\bibnamefont{Czurko}}, \bibnamefont{and}
  \bibinfo{author}{\bibfnamefont{G.}~\bibnamefont{Buzs{\'a}ki}},
  \bibinfo{journal}{Neuron} \textbf{\bibinfo{volume}{21}}, \bibinfo{pages}{179}
  (\bibinfo{year}{1998}).

\bibitem[{\citenamefont{Barth{\'o} et~al.}(2004)\citenamefont{Barth{\'o},
  Hirase, Monconduit, Zugaro, Harris, and
  Buzs{\'a}ki}}]{bartho2004characterization}
\bibinfo{author}{\bibfnamefont{P.}~\bibnamefont{Barth{\'o}}},
  \bibinfo{author}{\bibfnamefont{H.}~\bibnamefont{Hirase}},
  \bibinfo{author}{\bibfnamefont{L.}~\bibnamefont{Monconduit}},
  \bibinfo{author}{\bibfnamefont{M.}~\bibnamefont{Zugaro}},
  \bibinfo{author}{\bibfnamefont{K.~D.} \bibnamefont{Harris}},
  \bibnamefont{and}
  \bibinfo{author}{\bibfnamefont{G.}~\bibnamefont{Buzs{\'a}ki}},
  \bibinfo{journal}{Journal of neurophysiology} \textbf{\bibinfo{volume}{92}},
  \bibinfo{pages}{600} (\bibinfo{year}{2004}).

\bibitem[{\citenamefont{T{\"u}rker and Powers}(2001)}]{turker2001effects}
\bibinfo{author}{\bibfnamefont{K.}~\bibnamefont{T{\"u}rker}} \bibnamefont{and}
  \bibinfo{author}{\bibfnamefont{R.}~\bibnamefont{Powers}},
  \bibinfo{journal}{Journal of neurophysiology} \textbf{\bibinfo{volume}{86}},
  \bibinfo{pages}{2807} (\bibinfo{year}{2001}).

\bibitem[{\citenamefont{T{\"u}rker and Powers}(2004)}]{turker2004effects}
\bibinfo{author}{\bibfnamefont{K.}~\bibnamefont{T{\"u}rker}} \bibnamefont{and}
  \bibinfo{author}{\bibfnamefont{R.}~\bibnamefont{Powers}},
  \bibinfo{journal}{The Journal of Physiology} \textbf{\bibinfo{volume}{541}},
  \bibinfo{pages}{245} (\bibinfo{year}{2004}).

\bibitem[{\citenamefont{De~La~Rocha et~al.}(2007)\citenamefont{De~La~Rocha,
  Doiron, Eric Shea-Brown, and Reyes}}]{de2007correlation}
\bibinfo{author}{\bibfnamefont{J.}~\bibnamefont{De~La~Rocha}},
  \bibinfo{author}{\bibfnamefont{B.}~\bibnamefont{Doiron}},
  \bibinfo{author}{\bibfnamefont{K.~J.} \bibnamefont{Eric Shea-Brown}},
  \bibnamefont{and} \bibinfo{author}{\bibfnamefont{A.}~\bibnamefont{Reyes}},
  \bibinfo{journal}{Nature} \textbf{\bibinfo{volume}{448}},
  \bibinfo{pages}{802} (\bibinfo{year}{2007}).

\bibitem[{\citenamefont{Knight}(1972)}]{knight1972relationship}
\bibinfo{author}{\bibfnamefont{B.~W.} \bibnamefont{Knight}},
  \bibinfo{journal}{The Journal of general physiology}
  \textbf{\bibinfo{volume}{59}}, \bibinfo{pages}{767} (\bibinfo{year}{1972}).

\bibitem[{\citenamefont{Troyer and Miller}(1997)}]{troyer1997physiological}
\bibinfo{author}{\bibfnamefont{T.~W.} \bibnamefont{Troyer}} \bibnamefont{and}
  \bibinfo{author}{\bibfnamefont{K.~D.} \bibnamefont{Miller}},
  \bibinfo{journal}{Neural Computation} \textbf{\bibinfo{volume}{9}},
  \bibinfo{pages}{971} (\bibinfo{year}{1997}).

\bibitem[{\citenamefont{Tchumatchenko
  et~al.}(2010{\natexlab{b}})\citenamefont{Tchumatchenko, Geisel, Volgushev,
  and Wolf}}]{tchumatchenko2010signatures}
\bibinfo{author}{\bibfnamefont{T.}~\bibnamefont{Tchumatchenko}},
  \bibinfo{author}{\bibfnamefont{T.}~\bibnamefont{Geisel}},
  \bibinfo{author}{\bibfnamefont{M.}~\bibnamefont{Volgushev}},
  \bibnamefont{and} \bibinfo{author}{\bibfnamefont{F.}~\bibnamefont{Wolf}},
  \bibinfo{journal}{Frontiers in computational neuroscience}
  \textbf{\bibinfo{volume}{4}} (\bibinfo{year}{2010}{\natexlab{b}}).

\bibitem[{\citenamefont{Song et~al.}(2000)\citenamefont{Song, Miller, Abbott
  et~al.}}]{song2000competitive}
\bibinfo{author}{\bibfnamefont{S.}~\bibnamefont{Song}},
  \bibinfo{author}{\bibfnamefont{K.~D.} \bibnamefont{Miller}},
  \bibinfo{author}{\bibfnamefont{L.~F.} \bibnamefont{Abbott}},
  \bibnamefont{et~al.}, \bibinfo{journal}{Nature neuroscience}
  \textbf{\bibinfo{volume}{3}}, \bibinfo{pages}{919} (\bibinfo{year}{2000}).

\bibitem[{\citenamefont{Takahashi et~al.}(2009)\citenamefont{Takahashi, Kori,
  and Masuda}}]{takahashi2009self}
\bibinfo{author}{\bibfnamefont{Y.~K.} \bibnamefont{Takahashi}},
  \bibinfo{author}{\bibfnamefont{H.}~\bibnamefont{Kori}}, \bibnamefont{and}
  \bibinfo{author}{\bibfnamefont{N.}~\bibnamefont{Masuda}},
  \bibinfo{journal}{Physical Review E} \textbf{\bibinfo{volume}{79}},
  \bibinfo{pages}{051904} (\bibinfo{year}{2009}).

\bibitem[{\citenamefont{Bayati and Valizadeh}(2012)}]{bayati2012effect}
\bibinfo{author}{\bibfnamefont{M.}~\bibnamefont{Bayati}} \bibnamefont{and}
  \bibinfo{author}{\bibfnamefont{A.}~\bibnamefont{Valizadeh}},
  \bibinfo{journal}{Physical Review E} \textbf{\bibinfo{volume}{86}},
  \bibinfo{pages}{011925} (\bibinfo{year}{2012}).

\bibitem[{\citenamefont{Wang et~al.}(2012)\citenamefont{Wang, Chandrasekaran,
  Fernandez, White, and Canavier}}]{wang2012short}
\bibinfo{author}{\bibfnamefont{S.}~\bibnamefont{Wang}},
  \bibinfo{author}{\bibfnamefont{L.}~\bibnamefont{Chandrasekaran}},
  \bibinfo{author}{\bibfnamefont{F.~R.} \bibnamefont{Fernandez}},
  \bibinfo{author}{\bibfnamefont{J.~A.} \bibnamefont{White}}, \bibnamefont{and}
  \bibinfo{author}{\bibfnamefont{C.~C.} \bibnamefont{Canavier}},
  \bibinfo{journal}{PLoS computational biology} \textbf{\bibinfo{volume}{8}},
  \bibinfo{pages}{e1002306} (\bibinfo{year}{2012}).

\bibitem[{\citenamefont{Woodman and Canavier}(2011)}]{woodman2011effects}
\bibinfo{author}{\bibfnamefont{M.~M.} \bibnamefont{Woodman}} \bibnamefont{and}
  \bibinfo{author}{\bibfnamefont{C.~C.} \bibnamefont{Canavier}},
  \bibinfo{journal}{Journal of computational neuroscience}
  \textbf{\bibinfo{volume}{31}}, \bibinfo{pages}{401} (\bibinfo{year}{2011}).

\bibitem[{\citenamefont{Kuramoto}(1991)}]{kuramoto1991collective}
\bibinfo{author}{\bibfnamefont{Y.}~\bibnamefont{Kuramoto}},
  \bibinfo{journal}{Physica D: Nonlinear Phenomena}
  \textbf{\bibinfo{volume}{50}}, \bibinfo{pages}{15} (\bibinfo{year}{1991}).

\bibitem[{\citenamefont{Strogatz and Mirollo}(1991)}]{strogatz1991stability}
\bibinfo{author}{\bibfnamefont{S.~H.} \bibnamefont{Strogatz}} \bibnamefont{and}
  \bibinfo{author}{\bibfnamefont{R.~E.} \bibnamefont{Mirollo}},
  \bibinfo{journal}{Journal of Statistical Physics}
  \textbf{\bibinfo{volume}{63}}, \bibinfo{pages}{613} (\bibinfo{year}{1991}).

\bibitem[{\citenamefont{Abbott and van
  Vreeswijk}(1993)}]{abbott1993asynchronous}
\bibinfo{author}{\bibfnamefont{L.}~\bibnamefont{Abbott}} \bibnamefont{and}
  \bibinfo{author}{\bibfnamefont{C.}~\bibnamefont{van Vreeswijk}},
  \bibinfo{journal}{Physical Review E} \textbf{\bibinfo{volume}{48}},
  \bibinfo{pages}{1483} (\bibinfo{year}{1993}).

\bibitem[{\citenamefont{Ostojic et~al.}(2009)\citenamefont{Ostojic, Brunel, and
  Hakim}}]{ostojic2009connectivity}
\bibinfo{author}{\bibfnamefont{S.}~\bibnamefont{Ostojic}},
  \bibinfo{author}{\bibfnamefont{N.}~\bibnamefont{Brunel}}, \bibnamefont{and}
  \bibinfo{author}{\bibfnamefont{V.}~\bibnamefont{Hakim}},
  \bibinfo{journal}{The Journal of Neuroscience} \textbf{\bibinfo{volume}{29}},
  \bibinfo{pages}{10234} (\bibinfo{year}{2009}).

\bibitem[{\citenamefont{Babadi and Abbott}(2010)}]{babadi2010intrinsic}
\bibinfo{author}{\bibfnamefont{B.}~\bibnamefont{Babadi}} \bibnamefont{and}
  \bibinfo{author}{\bibfnamefont{L.~F.} \bibnamefont{Abbott}},
  \bibinfo{journal}{PLoS computational biology} \textbf{\bibinfo{volume}{6}},
  \bibinfo{pages}{e1000961} (\bibinfo{year}{2010}).

\bibitem[{\citenamefont{Gal{\'a}n et~al.}(2008)\citenamefont{Gal{\'a}n,
  Ermentrout, and Urban}}]{galan2008optimal}
\bibinfo{author}{\bibfnamefont{R.~F.} \bibnamefont{Gal{\'a}n}},
  \bibinfo{author}{\bibfnamefont{G.~B.} \bibnamefont{Ermentrout}},
  \bibnamefont{and} \bibinfo{author}{\bibfnamefont{N.~N.} \bibnamefont{Urban}},
  \bibinfo{journal}{Journal of neurophysiology} \textbf{\bibinfo{volume}{99}},
  \bibinfo{pages}{277} (\bibinfo{year}{2008}).

\bibitem[{\citenamefont{Abouzeid and Ermentrout}(2009)}]{abouzeid2009type}
\bibinfo{author}{\bibfnamefont{A.}~\bibnamefont{Abouzeid}} \bibnamefont{and}
  \bibinfo{author}{\bibfnamefont{B.}~\bibnamefont{Ermentrout}},
  \bibinfo{journal}{Physical Review E} \textbf{\bibinfo{volume}{80}},
  \bibinfo{pages}{011911} (\bibinfo{year}{2009}).

\bibitem[{\citenamefont{Barreiro et~al.}(2010)\citenamefont{Barreiro,
  Shea-Brown, and Thilo}}]{barreiro2010time}
\bibinfo{author}{\bibfnamefont{A.~K.} \bibnamefont{Barreiro}},
  \bibinfo{author}{\bibfnamefont{E.}~\bibnamefont{Shea-Brown}},
  \bibnamefont{and} \bibinfo{author}{\bibfnamefont{E.~L.} \bibnamefont{Thilo}},
  \bibinfo{journal}{Physical Review E} \textbf{\bibinfo{volume}{81}},
  \bibinfo{pages}{011916} (\bibinfo{year}{2010}).

\bibitem[{\citenamefont{Abouzeid and
  Ermentrout}(2011)}]{abouzeid2011correlation}
\bibinfo{author}{\bibfnamefont{A.}~\bibnamefont{Abouzeid}} \bibnamefont{and}
  \bibinfo{author}{\bibfnamefont{B.}~\bibnamefont{Ermentrout}},
  \bibinfo{journal}{Physical Review E} \textbf{\bibinfo{volume}{84}},
  \bibinfo{pages}{061914} (\bibinfo{year}{2011}).

\end{references}
\end{document}